\documentclass[conference]{IEEEtran}

\pdfoutput=1

\ifCLASSINFOpdf
\else
\fi

\usepackage{microtype}
\usepackage{graphicx}
\usepackage{multirow}
\usepackage{xcolor}
\usepackage[hidelinks]{hyperref}
\usepackage{tabularray}
\usepackage{tabularx}
\usepackage{array}
\usepackage{multirow}
\usepackage{tikz}
\tikzset{venn circle/.style={draw=gray,text opacity=1,fill opacity=0.25,circle,minimum width=10cm,fill=#1,line width=2pt}}
\tikzset{label/.style={text width=1.5cm,font=\large\sffamily}}

\usepackage[detect-weight=true,group-separator={,},
            output-decimal-marker={.}]{siunitx}
\makeatletter
\newcommand{\textlabel}[2]{%
	\protected@edef\@currentlabel{#1}
	\phantomsection
	\label{#2}
}
\makeatother

\begin{document}
\title{Towards Crowd-Based Requirements Engineering for Digital Farming (CrowdRE4DF)}

\author{
\IEEEauthorblockN{Eduard C. Groen\IEEEauthorrefmark{1}\IEEEauthorrefmark{2}, Kazi Rezoanur Rahman\IEEEauthorrefmark{3}, Nikita Narsinghani\IEEEauthorrefmark{3}, Joerg Doerr\IEEEauthorrefmark{1}\IEEEauthorrefmark{3}}

\IEEEauthorblockA{\IEEEauthorrefmark{1}Fraunhofer IESE, Germany, \{eduard.groen, joerg.doerr\}@iese.fraunhofer.de}

\IEEEauthorblockA{\IEEEauthorrefmark{2}Department of Information and Computing Sciences, Utrecht University, Netherlands}

\IEEEauthorblockA{\IEEEauthorrefmark{3}Department of Computer Science, University of Kaiserslautern-Landau (RPTU), Germany}
}

\maketitle

\begin{abstract}
The farming domain has seen a tremendous shift towards digital solutions. However, capturing farmers' requirements regarding Digital Farming (DF) technology remains a difficult task due to domain-specific challenges. 
%
Farmers form a diverse and international crowd of practitioners who use a common pool of agricultural products and services, which means we can consider the possibility of applying Crowd-based Requirements Engineering (CrowdRE) for DF: CrowdRE4DF. 
%
We found that online user feedback in this domain is limited, necessitating a way of capturing user feedback from farmers in situ. Our solution, the \textit{Farmers' Voice} application, uses speech-to-text, Machine Learning (ML), and Web 2.0 technology. 
%
A preliminary evaluation with five farmers showed good technology acceptance, and accurate transcription and ML analysis even in noisy farm settings. Our findings help to drive the development of DF technology through in-situ requirements elicitation.
\end{abstract}


\IEEEpeerreviewmaketitle

\section{Introduction} \label{sec:intro}

The agricultural revolution has brought significant transformations to the domain. The introduction of guidance and sensing systems, telematics, and data management has enabled precision farming, while the green revolution has helped to achieve high productivity gains~\cite{Lejon15}. More recently, smart farming has emerged, where physical products are supplemented by non-physical services in an agricultural ecosystem, which may include unmanned autonomous robotic and AI-based decision-making systems~\cite{Doerr21}. Using remote sensing, GPS, weather forecasting, and other technologies, farmers can make more informed decisions, automate agricultural work, make better predictions, and make their business processes more efficient. A collective term for these software-driven innovations is \textit{Digital Farming} (DF)~\cite{Doerr22}. Although the adoption and acceptance of new technology among farmers remains a limiting factor regarding the innovation potential in DF, development companies in the agricultural sector increasingly have a need to consider feedback from farmers when iteratively improving their (hardware-based) DF equipment and (software-based) DF applications. Farmers can provide contextual information, discuss their ideas, needs, challenges, and requirements, provide feedback on the actual systems they use, provide data for data-driven farming practices, and are a source of other relevant knowledge. As a result, Requirements Engineering (RE) can play an important role in providing solutions for eliciting such feedback from farmers about the technology they use, and in helping development companies improve and innovate their DF products.

Passive data collection methods, typically based on telematics data, are already in use, but these fall short in obtaining context-specific requirements from the farming community. It is likely that many requirements will be missed, which in turn may result in issues ranging from lower adoption rates of new technologies, wasted resources, inefficient solutions, and sustainability failures to complete market failure. Therefore, this knowledge needs to be augmented by feedback directly obtained from farmers. Farming is an international occupation performed by a diverse group of practitioners who use a common pool of products and services, so farmers can be considered a \textit{crowd}~\cite{Groen17ieee}. For this reason, it is worthwhile considering whether \textit{Crowd-based RE} (CrowdRE)~\cite{Groen17ieee} approaches are suitable to support farmers in communicating their feedback and development companies in using this feedback to further improve their products.

\textit{Research questions:}
We investigate the application of CrowdRE for DF (CrowdRE4DF) by considering its unique domain challenges and designing appropriate solutions through two consecutive studies with associated research questions: 

\begin{itemize}
    \item{\textbf{RQ1.}} What RE-relevant information is contained in online user feedback about DF products?
    \item{\textbf{RQ2.}} How can user feedback from farmers be successfully captured?
\end{itemize}

\textit{Research contributions:}
With this work, we make the following contributions:

\begin{enumerate}
    \item We introduce \textit{CrowdRE4DF} as a new subdiscipline of CrowdRE, and characterize it based on early empirical evidence. 
    \item We present a research prototype of the application \textit{Farmers' Voice} for eliciting speech-based user feedback from farmers, with which we performed an early validation.
    \item We provide the code for \textit{Farmers' Voice}, templates, and results from our studies as open artifacts in our online appendix~
\cite{appendix}.
\end{enumerate}

\textit{Paper structure:}
Section~\ref{sec:study1} presents our exploratory study into online user feedback, Section~\ref{sec:study2} describes the development and evaluation of our in-situ feedback elicitation tool \textit{Farmers' Voice}, and Section~\ref{sec:disc} discusses our RQs based on our findings. In Section~\ref{sec:rw}, we introduce the domain of DF and discuss relevant work on DF in RE, CrowdRE, and speech-to-text (S2T). Section~\ref{sec:plan} outlines our research plan, and Section~\ref{sec:concl} concludes this paper.

\section{Study I: Exploratory Analysis of User Feedback} \label{sec:study1}

The prevailing analysis approach in CrowdRE concerns online user feedback~\cite{Wang19}. Thus, our first study investigated the value of online user feedback on DF products. Online user feedback about smart products\textemdash such as smart equipment in DF\textemdash may differ from other types of applications~\cite{Groen17users}. User feedback may also be found in farming-specific online channels; this content may differ in terms of focus and language. Obtaining an understanding of these characteristics is helpful in determining how CrowdRE can best be tailored to DF.

\subsection{Methodology}

To establish the relevance of user feedback in DF for RE, we identified relevant sources of user reviews (Section~\ref{sec:study1-m1}), prepared the data (Section~\ref{sec:study1-m2}), and chose appropriate dimensions for classification (Section~\ref{sec:study1-m3}).

\subsubsection{Selection of Products and Data Sources} \label{sec:study1-m1}

DF products can be distinguished into (a)~\textit{DF equipment}, which includes sensors, tractors, and farming bots, as well as intelligent devices (e.g., Yara N-Sensor, Ecorobotix AVO), and (b)~\textit{DF applications}, which comprise software that is standalone or used in conjunction with hardware equipment. We determined that online user reviews about DF equipment are rare and typically not available through public channels. 
For DF applications, in addition to app stores (mainly Google Play), 
six websites provide publicly available user reviews: \href{https://www.farms.com/}{\underline{Farms.com}}, \href{https://www.fwi.co.uk/}{\underline{Farmers Weekly}}, \href{https://www.croplife.com/}{\underline{CropLife}}, \href{https://www.capterra.com/}{\underline{Capterra}}, \href{https://www.g2.com/}{\underline{G2.com}}, and \href{https://www.getapp.com/}{\underline{GetApp}}. We considered 24 farming applications (see Table~\ref{tab:apps}). Seven others were omitted, as they had received fewer than ten English-language reviews or their users are not farmers (e.g., crowdfarming applications). Two categories were the most prevalent: (1)~\textit{Precision agriculture} applications help to create farm maps, scout the farm, monitor installed sensors, etc., and (2)~\textit{farm management} applications provide insights and help to maximize yield and profits. 

\begin{table}[htb]
\centering
\caption{DF applications, associated number of unique reviews, and number of reviews in the sample in Study I.}
\label{tab:apps}
    \resizebox{\columnwidth}{!}{%
\begin{tabular}{lS[table-format=4.0]S[table-format=4.0]l}
\hline
\textbf{DF Application}    & \multicolumn{1}{c}{\begin{tabular}[c]{@{}c@{}}\textbf{Reviews}\end{tabular}} & \textbf{Sample} & \textbf{Category} \\
\hline
AgriApp                    & 994     & 100     & Farm Management        \\
AgriBus-NAVI               & 43      & 30      & GPS Guidance           \\
Agrio                      & 108     & 79      & Precision Agriculture  \\
Agrobase                   & 328     & 100     & Precision Agriculture  \\
Agroptima                  & 10      & 5       & Farm Management        \\
Atfarm                     & 14      & 12      & Precision Agriculture  \\
BharatAgri                 & 993     & 100     & Farm Management        \\
BoosterAGRO                & 13      & 4       & Farm Management        \\
Breeding Wheel             & 37      & 23      & Livestock Monitoring   \\
Climate FieldView          & 78      & 73      & Precision Agriculture  \\
CropX                      & 9       & 7       & Precision Agriculture  \\
FarmLogs                   & 148     & 100     & Farm Management        \\
Field Navigator            & 477     & 100     & GPS Guidance           \\
FieldBee                   & 95      & 74      & GPS Guidance           \\
GoHarvest                  & 61      & 43      & Vendor-Specific        \\
GPS Field Area Measure     & 1049    & 100     & GPS Guidance           \\
Landwirt.com Traktor Markt & 16      & 5       & Marketplace            \\
OneSoil Scouting           & 70      & 51      & Precision Agriculture  \\
Plantix                    & 999     & 100     & Precision Agriculture  \\
Soil Sampler               & 69      & 37      & Precision Agriculture  \\
tractorpool                & 12      & 7       & Marketplace            \\
xarvio Field Manager       & 23      & 18      & Precision Agriculture  \\
xarvio Scouting            & 522     & 100     & Precision Agriculture  \\
Yara CheckIT               & 112     & 67      & Precision Agriculture  \\
\hline
Total                      & 6280    & 1335     &     \\                  
\hline
\end{tabular}%
}
\end{table}

\subsubsection{Data Pre-Processing} \label{sec:study1-m2}

The online user reviews for all selected applications amounted to a dataset of 6,280 reviews, after removing 32 duplicates. 
Before sampling, we omitted reviews with $\leq$10 characters because these have little informational value (-1,941 reviews) as well as reviews that are written in a foreign language, contain spurious words, or consist predominantly of special characters or emojis (-703 reviews). Because the purpose of this study was to explore typical user feedback about DF applications, we created a bootstrapped random sample of 100 reviews per application to diminish the imbalance in the dataset, which resulted in a sample of 1,335 reviews (see Table~\ref{tab:apps}). 

\subsubsection{Review Classification} \label{sec:study1-m3}

The sample of reviews was manually annotated and reconciled by two authors. Because of the exploratory nature, 
we began with open coding over a sample of the longest reviews ($\geq$500 characters). 
We determined that the relevant content can be distinguished into the three perspectives listed below. Content that did not fit these three categories was earmarked for discussion, but this did not lead to any new classes. Because a review might address several topics, multiple classes could be assigned. Reviews without clear or relevant content were classified as ``None''.

\begin{enumerate}
    \item \textbf{System:} A feature, quality, or other technical aspect of the DF application, such as praise, criticism, or requests.
    \item \textbf{Operations:} Use cases regarding the value that the DF application brings to the day-to-day farming business.\footnote{Works that classified general-purpose applications used classes such as \textit{usage information} or \textit{functional suitability} (e.g.,~\cite{Groen17users,Anders24}) to annotate descriptions by users of how and in what context they use the application. Due to the specialized nature of farming, we assert that the non-technical and domain-specific descriptions of farming operations are markedly different.}
    \item \textbf{Customer Support:} Experiences receiving assistance or advice from the developer of the DF application, a specialist, or a dedicated community.  
\end{enumerate}

\subsection{Results} \label{sec:study1-r}

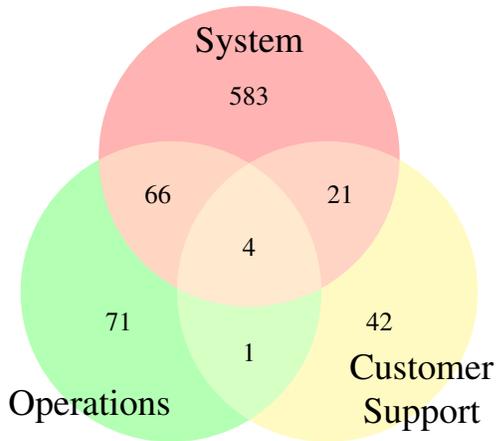
\begin{figure}[htb]
\centering
\begin{tikzpicture}
  \begin{scope}[blend group = soft light] 
    \fill[red!30!white]   ( 90:1.2) circle (2);
    \fill[green!30!white] (210:1.2) circle (2);
    \fill[yellow!30!white]  (330:1.2) circle (2);
  \end{scope}
  \node at ( 90:2)    {583}; 
  \node at ( 150:1.4) {66};  
  \node at ( 210:2)   {71};  
  \node at ( 270:1.4) {1};   
  \node at ( 330:2)   {42};  
  \node at ( 30:1.4)  {21};  
  \node               {4};   
  \node at (90:2.7) [font=\Large]  {System}; 
  \node at (225:3) [font=\Large]   {Operations}; 
  \node at (325:2.8) [font=\Large] {Customer}; 
  \node at (312:3) [font=\Large]   {\,\,\,\,\,\,\,Support}; 
\end{tikzpicture}
\caption{Distribution of the multi-label classifications.}
\label{fig:venn}
\end{figure}

Of the 1,335 reviews on DF applications, 547 (41.0\%) were considered irrelevant (i.e., ``none''). 
Thus, if the reviews with $\leq$10 characters were included, the share of relevant reviews would likely fall below 50\%. 
Figure~\ref{fig:venn} shows the distribution of the classes. In total, 674 reviews (85.5\%) address the \textit{system}, with 583 (74.0\%) of them doing so exclusively. These can be analyzed using existing classifiers in RE. A still considerable share of 18.0\% (9.0\% exclusively) relate to farming \textit{operations}, and 8.6\% (5.3\% exclusively) mention \textit{customer support}.

User reviews classified as \textit{system} most often request particular crops or livestock types, machines, languages, or geographical areas to be added. For example: \textit{``Major agriculture economy India is not present. What is the reason? Not able to obtain data from the Indian govt?''} Reviews also positively or negatively state whether expectations are met regarding the usefulness, accuracy, and currentness of map data; for example: \textit{``You say that every 3-5 days new satelite} [sic] \textit{photo comes, but how come I see only 2015 or even older photo instead? Idea quite good, but realisation not even close.''} Usability-related feedback was mostly positive, while other qualities only seemed to be addressed if the application did not start or crashed, failed to exchange data, was not responsive enough, used up too much bandwidth, or had login problems.

User reviews classified as \textit{operations} were typically positive and described the value of the application, which might reveal unique value propositions as well as unidentified potential for deployment in organic farming, for accurately documenting aspects, for taking measurements, etc. For example: \textit{``While it's true this app is designed more for row crop farms, I use it on our hay ground to keep track of rainfall.''}

User reviews classified as \textit{customer support} mostly praise the timeliness or quality of the responses from the development team, advisors, or online community. There are, however, also negative reviews about the responsiveness or politeness of customer service, such as: \textit{``If you tell them a problem, they will not listen to it for 48 hours. When the plants are completely damaged, their suggestion comes.''} Some reviews also praise the mode of communication (e.g., video calls) or ask how the developer can be contacted.

The strong overlap between \textit{operations} and \textit{system} partly stems from reviews praising the application's user-friendliness combined with a description of how a farmer uses the application, but more often a farmer motivates a request by explaining their particular situation. For example, a review combining a feature request related to a farm's location is: \textit{``I am still waiting} [sic] \textit{offline function. We work near border of republic. There are} [sic] \textit{no internet.''} Some reviews classified as \textit{system} implicitly reveal a property of the farm, but do not elaborate on the use case behind the request. 
For a small share of the reviews, we found patterns that indicated that these might be fake, for example when reviews reused sentences from other reviews verbatim, overly promoted the application, or described it as a game. These reviews typically did not contain useful information.

By further analyzing the contents, we found that it is possible to establish a vocabulary of farming-related terms in online user feedback that could be organized into eight categories: \textit{crops \& livestock} (e.g., grain, grow), \textit{property \& inventory} (e.g., farm, land), \textit{measurements} (e.g., cost, acre), \textit{weather \& irrigation} (e.g., rain, season), \textit{personnel \& equipment} (e.g., operator, truck), \textit{GPS guidance} (e.g., field, tracking), \textit{system} (e.g., farm management software, agronomy), and \textit{fertilizers \& pesticides} (e.g., chemicals, spray). Only a smaller subset of these terms are directly associated with DF applications or smart equipment, while the majority of the terms relate to aspects typically associated with farming. Only two user reviews that we found mentioned livestock, while many reviews used terms related to crop growing.

\subsection{Threats to Validity} \label{sec:study1-t}
We discuss the main validity threats according to Wohlin et al.~\cite{Wohlin00}. \textit{Construct validity:} 
We especially found applications related to crop farming, and cannot exclude that we missed relevant sources or applications (e.g., for livestock, forestry, or fishery;~\cite{vanderLinden21, Ferrari23}). We believe that our methodology helped to assure the integrity of our classifications and note that it served an exploratory purpose and did not seek to create a gold standard.
\textit{Internal validity:} Our findings are based on a small amount of data. 
We strove to be very inclusive about the applications we considered and tried to obtain as many farming-related reviews as possible. We did consider other applications but determined that the reviews did not suit our study. This included an application that received only reviews in German 
and a GPS navigation application that predominantly received general-purpose user feedback. 
\textit{External validity:} This work is only intended to characterize user feedback in the farming domain and the results are not very likely to extrapolate to other professional domains.

\subsection{Intermediate Discussion} \label{sec:study1-d}

Through this study, we gained insights into the feedback-giving behavior of farmers. The lack of DF feedback channels, the absence of reviews about DF equipment, and the low number of reviews each DF application receives (see Table~\ref{tab:apps}) together suggest that most farmers do not write user reviews. Software applications that users interact with typically receive many more reviews~\cite{Pagano13}. The number of reviews each application received can still be analyzed manually and does not warrant investing time and effort in an automated analysis pipeline~\cite{Groen18}, which in turn would forfeit the need to tailor CrowdRE to DF. Similarly, it is not worthwhile investing in a taxonomy on DF as we were only able to curate a total of about 6,000 reviews for this domain, while using the annotated data is prone to over-fitting. Nevertheless, a good proportion of the reviews is requirements-relevant~\cite{Sihag23}. Like domain-unspecific feedback, the feedback highlights functional and quality aspects of the system~\cite{Groen17users}, while valuable insights can also be gained from descriptions of the usage (i.e., farming operations) and the customer support received~\cite{Anders24}.
In Section~\ref{sec:disc}, we will discuss our key findings of Study I to answer RQ1. 

\section{Study II: Evaluation of a Speech-to-Text Feedback Solution} \label{sec:study2}

Study I revealed that due to the scant amount of user feedback farmers generate, automated user feedback analysis methods will not contribute much to a better understanding of this crowd of users. Discussions with domain experts allowed us to conclude that this is caused by the farmers' work being demanding. They lack the time to stand around typing a review on their smartphone and do not have much opportunity to type out a user review on a computer because they are not often at their desk. This is unlikely to change through motivational measures encouraging farmers to write more user reviews.

This finding gave rise to Study II, where we set out to design and evaluate a solution that takes into consideration the practical limitations faced by farmers while exploring other suitable ways of capturing user feedback from farmers. Considering their daily schedule, we dismissed any solutions that would require farmers to take the time to provide written feedback. When they are in the office, they are busy preparing or documenting the field work, and out in the field they cannot be reasonably expected to pause for a moment. However, just like they make phone calls during the day, for example while driving their tractor, they could record their feedback through speech in situ. This resulted in our research prototype \textit{Farmers' Voice}; a speech-to-text (S2T) and audio sentiment analysis (ASA) application that allows crowds of farmers to speak with one voice. We evaluated our implementation regarding its viability to help with feedback collection in terms of whether farmers would accept such a solution and whether it would work in a farm setting. The application and all resources necessary to replicate this study are available in our online appendix~\cite{appendix}.

\subsection{Analysis \& Conceptual Solution}

We found that existing S2T applications and smart voice assistants do not support integration into a pipeline, but determined that common S2T features include multi-language support, live transcription, recording control buttons (i.e., start, pause, stop, reset), offline mode, data privacy measures, user-friendliness, transcript editing, and tutorials.

To counteract the limited literature on DF and extend the research team's domain knowledge, we consulted with an agriculture economist who has been researching DF for four years and seasonally works on the family farm. He recommended postponing liaising with farmers until we could provide them with tangible results because they are often critical before adopting new technologies. 
His most important recommendations are:
(1)~It is crucial to farmers to be able to contact support personnel, e.g., technical contacts at their equipment dealer or agricultural cooperative and dedicated account managers at companies developing their business software. Farmers mostly call them, which further supports our choice of a speech-based solution. 
(2)~The likelihood that farmers will be able to provide user feedback while sitting on a tractor will only increase once automated steering systems become more common in modern tractors. Any solution should be able to handle the noise levels and work without an Internet connection. (3)~The use of smart voice assistants may be on the rise, but farmers often lack the time to interact with them. They are more likely to be willing to send a one-off voice message, especially when short feedback cycles show a benefit of investing their time.

We considered three types of \textit{stakeholders} for \textit{Farmers' Voice}: farmers, support personnel, and requirements engineers. In this paper, we focus on the \textit{farmers}, who will use the application to record and submit feedback about DF applications and DF equipment, such as (technical) problems, expectations, or ideas. Their willingness to use the application is central to its success. The feedback should then be sent to the \textit{support personnel}, e.g., technical contacts or account managers, who can respond directly in case of a technical problem and prioritize the feedback. The \textit{requirements engineers} can then analyze the user feedback for recurring patterns, needs, and ideas. 

\begin{table*}[t]
\centering
\caption{Requirements for the research prototype of \textit{Farmers' Voice}.}
\label{tab:requirements}
\begin{tabular}{l l}  
\hline
\textbf{ID} & \textbf{Requirement} \\
\hline
RQ01 & The application shall have a user interface that non-tech savvy users are able to operate. \\
RQ02 & The application shall allow the user to record feedback through speech and see the generated text-based transcript. \\
RQ03 & The application shall allow the user to submit their feedback in the form of raw audio clips or text-based transcripts. \\
RQ03 & The application shall be cross-platform to enable its use on as many platforms as possible, with an emphasis on mobile platforms. \\
RQ04 & The application must assure data privacy and include basic security features such as a login system and data removal policies. \\
RQ05 & The application shall process the transcribed text, at least regarding keywords, summaries, and sentiments. \\
RQ06 & The application shall allow the user to switch between German and English for both the interface and speech processing. \\
RQ07 & The application shall allow the user to download the transcript, processing results, and metrics as a report in an appropriate file format. \\
RQ08 & The application shall have an audio sentiment analysis module that processes audio recordings. \\
RQ09 & The application shall allow the user to submit both new and older recordings for audio sentiment analysis. \\
RQ10 & The application shall allow the user to toggle between the standard free-form feedback and a research-oriented baseline feedback module. \\
RQ11 & The application shall be deployed in an online hosting environment to make core components available in remote setups. \\
\hline
\end{tabular}
\end{table*}

Our analysis of S2T applications, our consultation with a domain expert, and iterative discussions among the authors resulted in eleven requirements for our research prototype, which are listed in Table~\ref{tab:requirements}. These were translated into use cases, which were then used for our interface design in {\href{https://www.justinmind.com/}{\underline{Justinmind}}. 
The S2T and ASA modules allow us to determine whether farmers prefer reviewing and submitting a text-based transcript or an audio recording augmented by sentiment and keyword indicators. The baseline feedback module was designed to store data differently and provide additional reporting functionalities necessary for evaluating the application's performance against predefined benchmarks. The user journey in the application as shown in Figure~\ref{fig:userjourney} is as follows: After logging in (screen 1), the user encounters a landing page where recordings can be made and managed in free-form feedback mode and S2T processing (2). The transcript is shown directly on-screen (3). In baseline mode, a report can be accessed (4\&5). The user can activate the ASA module in the tabs of the top navigation bar (6). The ASA module includes a media player for each recording and a menu to display the detected emotions (7), and the ability to upload recordings (8).

\subsection{Software Architecture \& Implementation}

We developed \textit{Farmers' Voice} in \href{https://code.visualstudio.com/}{\underline{Visual Studio Code}} following the \href{https://prettier.io/}{\underline{Prettier}} conventions, using \href{https://git-scm.com/}{\underline{Git}}, {\href{https://github.com/}{\underline{Github}}, and \href{https://huggingface.co/}{\underline{HuggingFace}} for version control, and \href{https://github.com/features/actions}{\underline{Github Actions}} for continuous integration and deployment. 
A high-level overview of the architecture and its components is shown in Figure \ref{fig:architecture}.
Users interact with the front-end through a web-based application based on the component-based architecture of \href{https://react.dev/}{\underline{ReactJS}}. We integrated Mozilla's \href{https://developer.mozilla.org/en-US/docs/Web/API/SpeechRecognition}{\underline{Web Speech API}} using the \href{https://legacy.reactjs.org/docs/hooks-intro.html}{\underline{react hook}} ``\href{https://www.npmjs.com/package/react-speech-recognition}{\underline{react-speech-recognition}}'', and reusable user interface components through the \href{https://mui.com/}{\underline{Material UI}} library. We designed \textit{Farmers' Voice} as a \href{https://web.dev/progressive-web-apps/}{\underline{progressive web application}} to ensure responsiveness, cross-platform adaptability (e.g., to different screen or window dimensions), and offline functionality to ensure that the web-based application will remain accessible in poor network conditions (e.g., out in the field). Sending the data to the model for processing over APIs does require an Internet connection. Switching languages and localization was achieved through the JavaScript-based \href{https://www.i18next.com/}{\underline{I18n}} internationalization framework. The front-end stack was deployed in \href{https://speechtotextresearch.web.app/}{\underline{Google Firebase}}.
The back-end provides data to the front-end components and handles post-processing. For natural-language processing, we use HuggingFace's pre-trained models, which are accessed using Python with \href{https://gradio.app/}{\underline{Gradio}}'s ``use via API'' endpoints deployed in \href{https://huggingface.co/spaces}{\underline{HuggingFace Spaces}}, through which we invoke Superb/wav2vec2-base-superb-er and distilbert-base-uncased-finetuned-sst-2-english for sentiment analysis, transformer3/keywordextractor for keyword extraction, and knkarthick/meeting\_summary for summarization. The \href{https://www.deepl.com/en/translator}{\underline{DeepL API}} deployed in \href{https://heroku.com/}{\underline{Heroku}} translated transcripts into English. 
We decided not to employ noise reduction algorithms in the research prototype because their consistency cannot be assured across devices; we tested in the evaluation whether there would be a need to implement such functionality during live transcription or when processing audio files.

\begin{figure*}[!t]
\centering
\includegraphics[width=\textwidth]{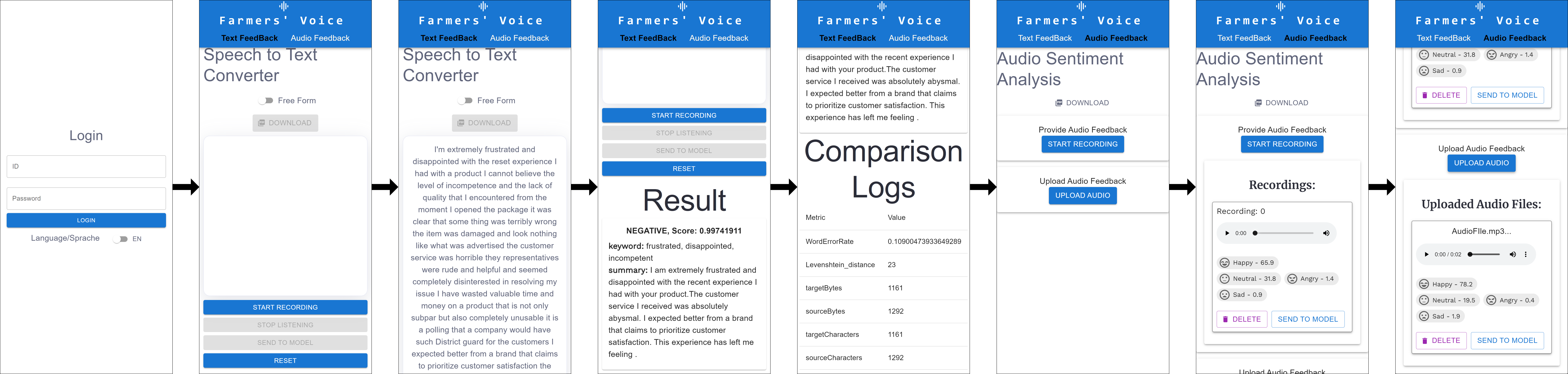}
\caption{Screenshots from the mobile layout of \textit{Farmers' Voice} showing a possible user journey.}\label{fig:userjourney}
\end{figure*}

\begin{figure}[hb]
\centering
  \includegraphics [width=\columnwidth]{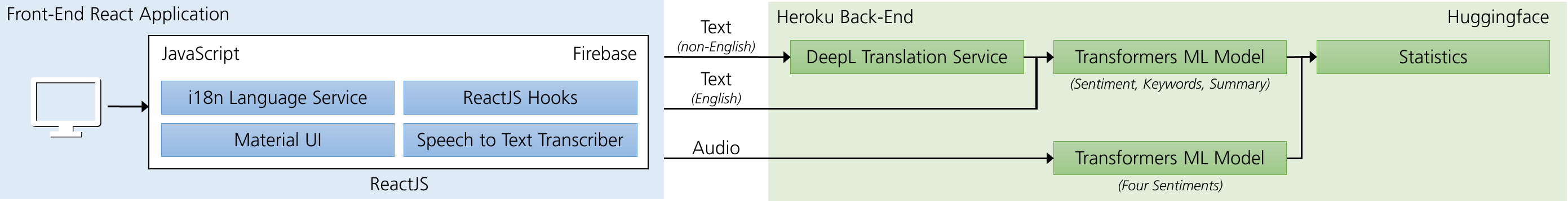}
  \caption{Overview of the architecture for \textit{Farmers' Voice}.}\label{fig:architecture}
\end{figure}

\subsection{Evaluation}
We evaluated \textit{Farmers' Voice} to measure its performance in realistic settings and determine the acceptance of our solution among prospective end users. We needed to make sure that the application will work in farm settings and would be used by farmers at all, before our research addresses the quality of its data processing of the user feedback content.

\subsubsection{Participants \& Instrumentation}
The evaluation was conducted at Hofgut Neum\"{u}hle, an associated farm specializing in research and training for livestock farming. Through its network, we solicited five farmers (one female, four male) from the German state of Rhineland-Palatinate who were available to participate on 2 August 2023: 
three experienced grain farmers (P1, P2, and P5), a novice vegetable farmer (P4), and an intermediate pasture farmer (P3). Regarding their technological proficiency, they were all very familiar with mobile applications and would feel comfortable using a speech-to-text application. Except for P4, they actively used technology professionally and had prior experience with voice assistants like Siri or Alexa. Developers of DF technology typically target farmers with these properties to market their solutions to. Most participants considered themselves rather good typists, except P3, who is a slow typist. However, P2, P3, and P5 stated that they would prefer giving feedback using speech, while P1 and P4 would additionally like to provide written feedback. 
The experiment was conducted using a OnePlus 7 Pro smartphone with Android 12 installed, which was tested prior to the experiment to prevent technical issues such as connectivity problems.

\subsubsection{Procedure}
The evaluation was conducted in three stages: (1)~briefing \& pre-evaluation survey (30 minutes); (2)~field experiment (75 minutes), and (3)~focus group \& debriefing (100 minutes). 
In the briefing, we introduced the participants to the research and familiarized them with the \textit{Farmers' Voice} application, its modules and workflow, and clarified questions. In the pre-evaluation survey, which is included in our online appendix~
\cite{appendix}, we elicited their demographics and expectations.

We performed a 2 (noise conditions) $\times$ 2 (feedback types) within-subjects field experiment. 
The same steps were performed in all scenarios by all five participants to ensure that any differences could be attributed to the change in the environment. Accordingly, the experiment consisted of 20 treatments.
Two noise conditions\textemdash office setting and field setting\textemdash simulated realistic scenarios in a farmer's day. The \textit{office setting} was located in a room at the farm and resembled a low-noise environment. 
For the \textit{field setting}, the farmer was seated on a tractor running stationary but with the power set to maximum (i.e., generating the loudest possible noise), to represent a high-noise environment. This allowed us to compare transcription accuracy. 
Two feedback types\textemdash baseline and free-form\textemdash determined the content. In the \textit{baseline feedback} condition, each participant was provided with the prepared baseline text included in our online appendix~\cite{appendix} to enable quantitative comparisons. They read this text aloud while the application was recording. 
In the \textit{free-form feedback} condition, the participants picked a topic, such as a new farming technique, a crop management strategy, or an experience with DF equipment, and freely discussed notable functionalities, ease of use, impact on farming operations, (dis)advantages, and suggestions for improvement.
In each condition, a recording was made separately using each module (S2T and ASA). This way, the farmers could get a feel for what it is like to use the application. Their acceptance of the proposed technology and perception of the application's usability were captured using a focus group to get feedback about the use of the application, satisfaction, perceived usefulness, challenges, and preferences.

\subsubsection{Data Analysis}

For data collection, \textit{Farmers' Voice} generated a report with several typical S2T metrics. The \textit{Word Error Rate (WER)}~\cite{Klakow02} measures the overall accuracy of the S2T model. It compares the number of errors (substitutions, deletions, and insertions) in a transcription to the total number of words against the baseline text. The \textit{Levenshtein Distance} or \textit{edit distance}~\cite{Levenshtein66} quantifies the differences between the original spoken text and the transcribed text to determine the accuracy of the transcription compared to the original speech recording. It measures the difference between two strings in terms of the minimum number of single-character edits required to transform one string into the other. We also compared the length of each transcription (\textit{target bytes and characters}) to the baseline text\textemdash 1,597 source bytes and 1,572 source characters\textemdash to determine the completeness of the transcription. 
We calculated the significance of the effect of noise conditions using a paired T-test, which was one-tailed because the field setting was hypothesized to perform worse than the office setting. Because five item pairs yield a low effect size, we set the significance level to $\alpha = 0.10$.

\subsubsection{Results}
The performance of \textit{Farmers' Voice} in both noise conditions is shown in Table~\ref{tab:voiceResults}. In the field setting, WER and Levenshtein distance are higher. Significance was not reached, but the trend suggests that a difference could be observed with more participants. The lower number of target bytes and characters in the field setting also reflects slightly greater difficulty in transcribing some participants' speech; e.g., with P2, the number of misinterpretations quadrupled in the field setting. However, the performance was not significantly worse in the field setting. This shows that the application is only slightly less accurate in a high-noise setting.

\begin{table}[htb]
    \centering
    \caption{Means and standard deviations (SD) for the settings and participants of the measured S2T performance in the baseline module, and the significance of the difference ($p$); the omitted values for difference equal the target values.}
    \label{tab:voiceResults}
    \resizebox{\columnwidth}{!}{%
\begin{tabular}{l|rr|rr|rrr}
  & \multicolumn{2}{c|}{\textbf{Office Setting}} & \multicolumn{2}{c|}{\textbf{Field Setting}} & \multicolumn{3}{c}{\textbf{Comparison}} \\
\textbf{Metric} & \multicolumn{1}{l}{\textbf{Mean}} & \multicolumn{1}{l|}{\textbf{SD}} & \multicolumn{1}{l}{\textbf{Mean}} & \multicolumn{1}{l|}{\textbf{SD}} & \multicolumn{1}{l}{\textbf{Mean}} & \multicolumn{1}{l}{\textbf{SD}} & \multicolumn{1}{l}{\textbf{\textit{p}}} \\ \hline
Word Error Rate      & 0.30   & 0.07  & 0.36   & 0.08  & 0.05   & 0.01  & 0.1109  \\
Levenshtein Distance & 67.2   & 15.1  & 79.4   & 17.2  & 12.20  & 2.12  & 0.1040  \\
Target Bytes         & 1422.8 & 105.7 & 1350.6 & 158.9 & -72.20 & 53.20 & 0.2180  \\
Byte Difference      & -174.2 & 105.7 & -246.4 &       &        &       &         \\
Target Characters    & 1398.8 & 104.9 & 1329.0 & 157.8 & -69.80 & 52.88 & 0.2242  \\
Character Difference & -173.2 & 104.9 & -243.0 &       &        &       &         \\ \hline
\end{tabular}%
}
\end{table}

The participants' acceptance of the application increased during the experiment. They were critical in the pre-evaluation survey, expecting an S2T application to capture their feedback accurately and reliably if they were to accept it. In the focus group, all were positive about the use of the application, with expectations exceeded for some. 
They found that the transcription worked intuitively and fast, and returned accurate feedback, with P1 saying: \textit{``During free-form feedback, I said many different words and I was impressed with the performance.''} They did notice that there were problems with common farming abbreviations like NIR and NDVI,\footnote{Near-Infrared; Normalized Difference Vegetation Index.} which should be supported to handle reports of complex topics. They also expressed concerns about the transcription handling heavy dialects, of which there are almost 250 in Germany alone. They recommended addressing both issues sooner rather than later.

The participants agreed that the application was comfortable to work with and that they could provide feedback with ease. This was even the case for P4, who had never used voice assistants. They would find it easier to provide feedback in this way than by contacting their dealer or other support personnel, but P5 noted that emotions can be better expressed through personal communication. All participants indicated that they preferred providing feedback verbally rather than typing it. As a slow typist, P3 found this to be both more convenient and a time-saver. This was in line with the expectations expressed in the survey. This evidence suggests that S2T can be an easier way for farmers to provide feedback. Four participants preferred the S2T module, while P3 preferred the audio sentiment analysis module because \textit{``one can draw better conclusions about the customer’s mood.''} The participants found that the ASA module was not very accurate in identifying emotions other than anger; it tended to label most feedback as either neutral or sad.

In the pre-evaluation survey, data privacy was a major concern. Trust increased during the experiment, but the participants agreed that many farmers may be concerned about how the data is used and who will receive it. They provided a compelling example: Farmers might fear that if feedback about their fertilizer use reaches federal or state authorities, they could use it to impose further regulations or restrictions. 

\subsection{Threats to Validity}

We discuss the main validity threats according to Wohlin et al.~\cite{Wohlin00}. 
\textit{Construct validity:} We strove to evaluate the most crucial features of \textit{Farmers' Voice}. It was not possible to balance the small number of participants across treatments or draw statistically significant conclusions about the conditions. To diminish instrumentation effects, we used a single smartphone without noise cancellation features. Android limits the microphone listening time of an app, which we circumvented by instructing the participants to avoid longer pauses. 
\textit{Internal validity:} To mitigate the Hawthorne effect during the experiment or the Rosenthal effect during the focus group due to the development and evaluation being performed by the same researcher, we included a neutral second experimenter and carefully prepared the guidelines, among other things to facilitate an atmosphere in which the participants felt comfortable to speak openly.
\textit{External validity:} We evaluated the application with professional farmers; a stakeholder group that is difficult to involve in experiments due to their limited availability. Despite their diversity in terms of gender, experience, technological affinity, and type of cultivation, we cannot exclude a selection bias towards farmers who (a)~are from the same region in Germany and (b)~might be more accepting of technological innovations. Regarding the latter, the participants were skeptical before the evaluation and were asked in the focus group to extrapolate whether they thought that other farmers would accept the application. Also, a critical attitude towards technology does not mean complete refusal to use a technology. The outcomes of our evaluation may be limited to the domain of farming.

\section{General Discussion} \label{sec:disc}

In this work, we set out to investigate whether CrowdRE can be applied to DF despite aggravating domain characteristics, and what tailoring may be necessary. 
The limited amount of research on DF required us to address many questions and assumptions along the way. 
Several challenges may inhibit the success of CrowdRE. Even more than in other professional fields, farmers have little opportunity to give feedback. Especially outside conditions for providing feedback might be unfavorable due to workload, weather, noise, Internet connectivity, etc. 
Many farmers are reluctant to adopt DF technologies, often out of insecurity over their technical literacy, English proficiency, or the trustworthiness of the application's privacy~\cite{Ofori20}. Adoption is greater among farmers with larger farms, better support infrastructures, and exposure to recommendations from peers. 
Privacy concerns and mistrust of governing bodies might also cause farmers to fear that their thoughts and ideas are misused.
The user feedback from farmers might also be harder to analyze automatically due to different languages and dialects, jargon, and abbreviations.

As to whether online user feedback from the farming domain contains RE-relevant information \textbf{(RQ1)}, we found too little feedback about DF equipment, some feedback on (software-based) DF products especially on Google Play, and few domain-specific or professional channels. RE-relevant information in the reviews is usually about the system, similar to online user feedback about general-purpose software applications, so it can be classified into features, qualities, or other RE-related dimensions by existing means (e.g.,~\cite{Kurtanovic17,Maalej15}).
Descriptions of how a DF application is used in farming operations may provide relevant insights to RE (cf.~\cite{Anders24}).
Although we can positively answer RQ1 based on the relevance of the content of online user feedback, DF applications are currently receiving too little feedback to allow for automated classification. 
Giving feedback might conflict with a farmer's daily schedule, which might be a unique challenge for applying CrowdRE in DF.

Investigating the question of how to capture user feedback from farmers in the best way \textbf{(RQ2)}, we adapted our strategy to their daily routine using \textit{Farmers' Voice}, an in-situ solution that allows spoken messages to be recorded, optionally transcribed, and submitted. 
It performed surprisingly well in a noisy tractor cabin even without the use of noise reduction technology. This means that giving feedback in a time and manner that is convenient for farmers, i.e., hands-free while driving a tractor, is a viable option. 
The farmers in our evaluation were open to using \textit{Farmers' Voice}. The experiment reduced their initial hesitation and met their expectations in terms of accuracy and usability. They did, in fact, prefer giving verbal feedback rather than written feedback and favored the S2T module over the ASA module. Accordingly, these technologies have the potential to successfully capture user feedback from farmers, thus to realize CrowdRE4DF.

\section{Related Work} \label{sec:rw}

Digital Farming (DF) refers to connected, knowledge-based agricultural production systems that utilize intelligent network and data management tools to automate sustainable processes in agriculture~\cite{CEMA17}. Its key components include sensors, Internet of Things devices, data analytics, autonomous robots and drones, and software applications. Value is created by connecting smart machines, with a huge emphasis on interoperability~\cite{Doerr23}. Farms are often challenged by the costs of implementing new technologies, weak and unstable Internet signals, or low bandwidth on the field. Older generations of farmers often depend on their digital native children for solving technical issues. Ofori et al.~\cite{Ofori20} found that, accordingly, farmers are quicker to adopt embodied-knowledge technologies (i.e., knowledge that is gained through direct experiences and interactions) than information-intensive technologies. 
Pfeiffer et al.~\cite{Pfeiffer2021} established that a positive attitude of German citizens towards farming and trust in farmers correlated with a positive attitude towards the perceived benefits of DF technologies, which include improved quality of life for farmers through reduced workloads, more environmentally-friendly production, and improved animal welfare and overall health. 

In RE, DF has received fairly limited attention. Blasch et al.~\cite{Blasch22} found that the adoption of Precision Farming technologies in Central Italy was moderated by initial investment costs, farm size, and economies of scale. Factors positively influencing adoption were financial support and networking and knowledge sharing among farmers, which demonstrates the important role of recommendations of new technologies from peers. 
Mannari et al.~\cite{Mannari23} found that class, goal, and process models were helpful for communication during requirements elicitation in a smart irrigation pilot study in Italy, even with non-technical stakeholders such as agronomists and farmers. 
Ferrari et al.~\cite{Ferrari23} established that farmers in remote mountain areas benefit from digital technology in terms of efficiency, reduced work, and strengthened social bonds, but that it can also form a distraction, limit problem-solving capability, and impair working conditions.

CrowdRE is an umbrella term for (semi-)automated approaches for collecting and analyzing information from a crowd to derive validated user requirements~\cite{Groen17ieee}. It presumes a crowd to be a sizable group of current or potential users with a common interest in a (software) product.
Collecting text-based feedback has received more attention than monitoring usage data~\cite{Khan19} and mainly involves user feedback analysis~\cite{Pagano13,Wang19}. Several mobile applications in RE elicit in-situ user feedback in the form of text and images, notably \textit{MyExperience}~\cite{Froehlich07}, \textit{iRequire}~\cite{Seyff10}, \textit{ConTexter}~\cite{Wehrmaker12}, \textit{AppEcho}~\cite{Seyff14}, and \textit{Rich Parking}~\cite{Wuest19}. 
Johnson et al.~\cite{Johnson20} performed the only CrowdRE study to date in the domain of farming by identifying CrowdRE challenges for software ecosystems for the case of \href{https://www.figured.com/}{\underline{Figured}}, a farming financial management system that is a partner application in the accounting ecosystem \href{https://www.xero.com/}{\underline{Xero}}. 
Farming development companies have an information need, but it is challenging to motivate farmers to provide user feedback.

We found only few demonstrations of S2T applied in RE. Soares et al.~\cite{Soares15} proposed the \textit{VoiceToModel} framework to improve the accessibility of the modeling process. 
With it, people with physical or visual impairments could create simpler goal, conceptual, and feature models by using voice commands. Their architecture demonstrates that such an application requires multiple external components to be integrated.
The pipeline by Maghilnan and Rajesh~\cite{Maghilnan17} demonstrates that speech-based sentiment analysis additionally requires supporting speech recognition and speaker recognition to distinguish between two or more speakers. For space reasons, we do not discuss the state-of-the-art approaches in S2T research here.

\section{Research Plan} \label{sec:plan}
Our speech-based feedback solution respects the farmers' daily schedules and has the potential to be accepted among farmers.
%
The participants in our evaluation were satisfied with the processing by the application. A next step is to collect actual feedback from farmers. Together with the stakeholder category of support personnel, we will seek to determine the quality of the feedback received. An open research question is whether farmers will prefer giving unstructured feedback or being interviewed by a conversational agent asking guiding questions such as (a)~``What do you think of the product?'' (i.e., \textit{system}), (b)~``How do you use the product?'' (i.e., \textit{operations}), and (c)~``How can we help you better?'' (i.e., \textit{customer support}). We also plan to improve the interaction, privacy, and transcript editing functionality of \textit{Farmers' Voice}. 

Giving a large crowd of farmers a voice may have a profound impact on their ability to steer decision-making processes of which they otherwise are only at the receiving end. One example are measures to make regions more resilient to the effects of climate change. We are preparing a research project in an area with a high geographic density of farmers that aims to not only passively collect requirements, but to specifically probe for their expertise in the co-creation of solutions. Such an influx of user feedback from farmers will provide us with the opportunity to design a robust CrowdRE4DF analysis pipeline that can achieve an accurate automated derivation of requirements from the user feedback: the central notion of CrowdRE~\cite{Groen17ieee}. Realizing such a pipeline is challenging because of the need to augment it with farming-related data that helps support the domain language and farmers' natural dialects. The vocabulary mentioned in Study I is a first step in this direction. Like other applications suggested~\cite{Seyff10, Wehrmaker12}, the pipeline may also need to support multi-modal feedback.

Offering speech-based feedback technology across systems while assuring cybersecurity reveals two limitations that can only be overcome if this is incorporated into domain standards such as the interface of the agricultural communication protocol ISOBUS~\cite{ISO17} or within the \href{https://gaia-x.eu/}{\underline{GAIA-X}} ecosystem. Farmers could then provide feedback directly through such a communication system. This would help assure data sovereignty and processing within a trustworthy environment. With sufficient DF technology such as sensors and Internet of Things devices in place, a context-aware system could also encourage farmers to provide feedback when they detect events such as malfunctions.

Although we will initially cater to farmers' needs, we recognize the potential for pivoting to domains and contexts with similar challenges. S2T technology could allow other stakeholders to give feedback in situ, such as factory workers operating machinery in noisy environments, drivers who should keep their hands on the steering wheel, or underground mine workers dealing with an unreliable communication network. Giving hands-free user feedback supports better inclusiveness of persons with an impaired ability to operate a device due to pain or physical disabilities, or who have trouble reading and writing due to visual impairments or illiteracy.

\section{Conclusion} \label{sec:concl}
In this paper, we analyzed the potential of applying CrowdRE in DF. There are unique challenges, which, among other things, cause only little online user feedback to be available. Enabling farmers to record feedback during their daily activities shows promise to help drive the development of DF technology. We have shown that farmers are ready to accept the idea and that our application provides accurate results even in noisy farming environments.  This gives DF technology developers direct access to farmers' feedback, while farmers can become the main drivers of technological advances in DF and improve their ability to produce food more sustainably.

\section*{Acknowledgments}
The authors would like to thank Daniel Eberz-Eder of the Dienstleistungszentrum L\"{a}ndlicher Raum Rheinhessen-Nahe-Hunsr\"{u}ck for arranging our contacts, Dr. Christian Koch of the Lehr- und Versuchsanstalt für Viehhaltung, Hofgut Neum\"{u}hle for providing his facilities, the farmers who generously donated their time to participate in our evaluation, Christoph Merscher for assisting with the experiment, and Sonnhild Namingha for proofreading this paper.

\ifCLASSOPTIONcaptionsoff
  \newpage
\fi

\bibliographystyle{IEEEtran}
\bibliography{IEEEabrv,references}

\end{document}